\documentclass[prd,twocolumn,showpacs,floatfix,superscriptaddress,nofootinbib]{revtex4}
\usepackage{graphicx}
\usepackage{epsfig}
\usepackage{bm}
\usepackage{amssymb}
\usepackage{float}
\usepackage{amsmath}
\usepackage{dcolumn}
\usepackage{cancel}
\usepackage[colorlinks]{hyperref}
\usepackage[usenames,dvipsnames]{color}
\hypersetup{
     breaklinks=true,
    pdfstartview={FitH},    
    colorlinks=true,       
    linkcolor=blue,          
    citecolor=red,        
    filecolor=magenta,      
    urlcolor=blue,           
    anchorcolor=green,      
    linktocpage=true
}

\def\doi{http://doi.org}
\def\arxiv{http://arxiv.org/abs}

\begin{document}

\title{Evading the  theoretical  no-go theorem for nonsingular bounces in 
Horndeski/Galileon cosmology}

\author{Shreya Banerjee }\email{shreya.banerjee@tifr.res.in}
\affiliation{Tata Institute of Fundamental Research, Homi Bhabha Road, Mumbai 400005, 
India}

\author{Yi-Fu Cai}
\email{yifucai@ustc.edu.cn}
\affiliation{CAS Key Laboratory for Research in Galaxies and Cosmology, Department of 
Astronomy, 
University of Science and Technology of China, Hefei 230026, China}
\affiliation{School of Astronomy and Space Science, University of Science and Technology 
of China, 
Hefei 230026, China}

\author{Emmanuel N. Saridakis}
\email{Emmanuel\_Saridakis@baylor.edu}
\affiliation{Department of Physics, National Technical University of Athens, Zografou
Campus GR 157 73, Athens, Greece}
\affiliation{Department of Astronomy, School of Physical Sciences, University of Science 
and Technology of China, Hefei 230026, P.R. China}


\begin{abstract}
We show that a nonsingular bounce, free of ghosts and gradient instabilities, can be 
realized in the framework of Horndeski or generalized Galileon cosmology. In particular, 
we first review that the theoretical {\it no-go} theorem, which states that the above is 
impossible, is based on two very strong assumptions, namely that a particular quantity 
cannot be discontinuous during the bounce, and that there is 
only one bounce. However, as we show in the present work, the first assumption 
 not only   can be violated in a general Horndeski/Galileon scenario, but also it is 
necessarily violated at the bounce point within the subclass of Horndeski/Galileon gravity 
in which $K(\phi,X)$ becomes zero at $X=0$. Additionally, concerning the second 
assumption, which is crucial in improved versions of the theorem which claim that even if 
a nonlinear free of pathologies can be realized it will lead to pathologies in the 
infinite past or infinite future, we show that if needed it can be evaded by considering 
cyclic cosmology, with an infinite sequence of  nonsingular bounces free of pathologies, 
which forbids the universe to reach the ``problematic'' regime at infinite past or 
infinite future. Finally, in order to make the analysis more 
transparent we provide explicit examples where nonsingular bounces without theoretical 
pathologies can be achieved.

\end{abstract}

\maketitle

\section{Introduction}

Nonsingular bouncing cosmologies may offer a potential solution to the problem of 
cosmological singularity \cite{Mukhanov:1991zn}. In particular, although inflation is 
considered to be a crucial part of the history of our universe \cite{inflation}, it is 
still accompanied by the above problem, since such a big bang singularity is unavoidable 
if inflation is driven by a scalar field in the framework of general relativity 
\cite{Borde:1993xh}. Hence, alongside the efforts to alleviate the 
initial singularity through quantum gravity effects, a significant amount of research 
directs towards its solution through the bounce realization. 

Bounce cosmology \cite{Novello:2008ra, Brandenberger:2012zb, Cai:2014bea, 
Battefeld:2014uga, Brandenberger:2016vhg, Cai:2016hea} can be realized by various modified 
gravity constructions \cite{Nojiri:2006ri, Capozziello:2011et, Cai:2015emx}, such as the 
Pre-Big-Bang \cite{Veneziano:1991ek} and the Ekpyrotic \cite{Khoury:2001wf, Khoury:2001bz} 
scenarios, higher-order gravity \cite{Tirtho1, Nojiri:2013ru}, $f(R)$ gravity 
\cite{Bamba:2013fha, Nojiri:2014zqa, Pavlovic:2017umo}, $f(T)$ 
gravity \cite{Cai:2011tc}, massive gravity \cite{Cai:2012ag}, braneworld models 
\cite{Shtanov:2002mb, Saridakis:2007cf}, non-relativistic gravity \cite{Cai:2009in, 
Saridakis:2009bv}, loop quantum cosmology \cite{Bojowald:2001xe, Cai:2014zga, 
Odintsov:2015uca}, Lagrange modified gravity \cite{Cai:2010zma} etc. Alternatively, 
nonsingular bouncing cosmology may be studied through the 
application of effective field theory techniques,  and the introduction of matter sectors 
that 
violate the null energy condition \cite{Cai:2007qw, Cai:2008qw, Cai:2009zp, 
Nojiri:2015fia}, or of 
non-conventional mixing terms \cite{Saridakis:2009jq, Saridakis:2009uk}. Such 
constructions can additionally provide an explanation for the scale invariant power 
spectrum \cite{Wands:1998yp, Finelli:2001sr} and 
moderate non-Gaussianities \cite{Cai:2009fn, Li:2016xjb}. In summary, bouncing cosmology 
may be considered as a potential alternative to the big bang one. 

A  general class of gravitational modification are the so-called galileon theories 
\cite{Nicolis:2008in, Deffayet:2009wt, Deffayet:2009mn, DeFelice:2011bh}, which are a 
re-discovery of the general scalar-tensor theory constructed by Horndeski under the 
requirement of maintaining the equations of motion second-ordered \cite{Horndeski:1974wa}. 
Application of the Horndeski/Galileon theory at a 
cosmological framework  proves to be very interesting and thus it has been investigated in 
detail in the literature. In particular, one can study the late-time acceleration 
\cite{Silva:2009km, Gannouji:2010au, DeFelice:2010pv, Tretyakov:2012zz, Leon:2012mt}, 
inflation \cite{Creminelli:2010ba, 
Kobayashi:2010cm, Ohashi:2012wf} and non-Gaussianities \cite{Mizuno:2010ag, Gao:2011qe, 
RenauxPetel:2011uk}, cosmological perturbations \cite{Kobayashi:2009wr, DeFelice:2010as, 
Barreira:2012kk}, or use observational data to extract constraints on various sub-classes 
of the theory \cite{Ali:2010gr, Iorio:2012pv, Appleby:2012ba}.

One interesting feature of Horndeski/Galileon theories is that they offer the framework 
for the realization of bouncing cosmology. In particular, one can obtain bouncing 
solutions in various sub-classes of the theory, describing both the background evolution 
as well as the generation of perturbations \cite{Qiu:2011cy, Easson:2011zy, Cai:2012va, 
Osipov:2013ssa, Qiu:2013eoa, Battarra:2014tga, Qiu:2015nha, Banerjee:2016hom, 
Ijjas:2016tpn, Ijjas:2016vtq, Ijjas:2017pei, Saridakis:2018fth}. Despite the success of 
Horndeski/Galileon theories in generating nonsingular bouncing solutions, there is a 
discussion on whether these solutions are stable. In particular, in 
\cite{Kobayashi:2016xpl, Akama:2017jsa, Kolevatov:2016ppi, Kolevatov:2017voe} the 
authors presented a  theoretical  {\it no-go} theorem stating that nonsingular models 
with flat spatial sections suffer in general from gradient instabilities or pathologies in 
the tensor sector. The proof of this theorem 
is based on two strong assumptions, namely that a specific non-observable quantity 
related to the 
tensor perturbation remains finite at the bounce point, and that there is only one 
bounce. However, this 
is not the general case, and indeed one can show that in successful and stable bounces 
the above assumption(s) are violated. Hence, the above theorem can be evaded and stable 
nonsingular bounces can be safely realized in the framework of Horndeski/Galileon 
cosmology. For instance, with the correspondence 
between the effective field theory (EFT) formalism and Horndeski/Generalized Galileon 
theories made in \cite{Gleyzes:2013ooa}, one may avoid this issue in bounce cosmology by 
modifying the dispersion relation for cosmological perturbations with the help of certain 
EFT operators \cite{Cai:2016thi, Cai:2017tku, Cai:2017dyi}.

In the following we explicitly show how the  theoretical  {\it 
no-go} theorem on nonsingular bounces in Horndeski/Galileon cosmology can be evaded.
We mention here that there is another {\it no-go} theorem from the observational 
perspective, which indicates that the parameter space for single-field nonsingular 
bounces is extremely limited due to the severe tension between tensor-to-scalar ratio and 
primordial non-gaussianity \cite{Quintin:2015rta, Li:2016xjb} (which in turn needs 
additional mechanisms to amplify the scalar perturbations \cite{Fertig:2016czu}). In the 
present work we refer only to the theoretical no-go theorem, namely our goal 
is to show that there is not a ``theoretical no-go theorem'', in the sense of a 
mathematically proven theorem of general validity, that forbids a non-singular bounce, and 
not to construct a bounce in perfect agreement with every observational requirement (which 
would require the thorough incorporation of background (SNIa, BAO, CMB shift parameter, 
$H_0$ measurements, etc) as well as perturbation ($f\sigma_8$) related data). Hence, even 
if a nonsingular bounce is difficult to be constructed from the observational point of 
view, it is not mathematically impossible.

The 
plan of the manuscript is as follows: In Section \ref{nogotheorem} we review the 
 theoretical  {\it no-go} theorem, mentioning the assumptions on which it is based. In 
Section \ref{Evading} we show that the aforementioned theorem is based on two strong 
assumption which for general sub-classes of the theory can be violated, and thus 
offering a safe evading of the theorem. Additionally, we provide   explicit examples 
where  nonsingular bounces free of ghost and gradient instabilities can be realized in 
Horndeski/Galileon cosmology. Finally, in Section \ref{Conclusions} we summarize the 
obtained results.

\section{The theoretical {\it no-go} theorem}
\label{nogotheorem}

In this section we review the  theoretical  {\it no-go} theorem which under specific 
assumptions states that nonsingular bounces in Horndeski/Galileon cosmology exhibit 
gradient instabilities or pathologies, following \cite{Kobayashi:2016xpl, Akama:2017jsa}. 

We start by presenting Horndeski, or equivalently the generalized Galileon theory, and 
its 
cosmological application. The corresponding action  is given by \cite{DeFelice:2011bh}
\begin{equation}
 S=\int d^{4}x\sqrt{-g}\,\sum_{i=2}^{5}{\cal L}_{i}\,,\label{Hor_L}
\end{equation} 
with
\begin{eqnarray}
&&
\!\!\!\!\!\!\!\!\!\!\!\!\!\!\!
{\cal L}_{2} = K(\phi,X),\label{eachlag2}\\
&&
\!\!\!\!\!\!\!\!\!\!\!\!\!\!\!
{\cal L}_{3} = -G_{3}(\phi,X)\Box\phi,\\
&&
\!\!\!\!\!\!\!\!\!\!\!\!\!\!\!
{\cal L}_{4} = G_{4}(\phi,X)\,
R+G_{4,X}\,[(\Box\phi)^{2}\!-\!(\nabla_{\mu}\nabla_{\nu}\phi)\,(\nabla^{\mu}
\nabla^{\nu}\phi)],
\end{eqnarray}
\begin{eqnarray}
&&
\!\!\!\!\!\!\!\!\!\!\!\!\!\!\!
{\cal L}_{5} = G_{5}(\phi,X)\,
G_{\mu\nu}\,(\nabla^{\mu}\nabla^{\nu}\phi)\,\nonumber\\
&&   -\frac{1}{6}\,
G_{5,X}\,[(\Box\phi)^{3}-3(\Box\phi)\,(\nabla_{\mu}\nabla_{\nu}\phi)\,
(\nabla^{\mu}\nabla^{\nu}\phi)\nonumber\\
&&\ \ \
\ \ \  
\ \ \ \  
\ \ \ \, 
+2(\nabla^{\mu}\nabla_{\alpha}\phi)\,(\nabla^
{\alpha}\nabla_{\beta}\phi)\,(\nabla^{\beta}\nabla_{\mu}\phi)],\label{
eachlag5}
\end{eqnarray}
with $R$ the Ricci scalar and $G_{\mu\nu}$ the Einstein tensor, and where we have set the 
Planck 
mass and the gravitational constant to $M_{pl}^{-2}\equiv 8\pi G=1$ for simplicity. The 
functions 
$K$ and $G_{i}$ ($i=3,4,5$) depend on the scalar field $\phi$ and its kinetic energy 
$X=-\partial^{\mu}\phi\partial_{\mu}\phi/2$, and moreover  $G_{i,X}\equiv\partial 
G_{i}/\partial X$ and $G_{i,\phi}
\equiv\partial G_{i}/\partial\phi$.

Applying the above theory in a cosmological framework, namely imposing a flat 
Friedmann-Robertson-
Walker (FRW) background geometry with metric 
\begin{eqnarray}
 ds^{2}=-dt^{2}+a^{2}(t)\delta_{ij}dx^{i}dx^{j},
\label{metric}
\end{eqnarray}
with $t$ the cosmic time, $x^i$ the comoving spatial coordinates, and $a(t)$ is the scale 
factor, 
one can extract the Friedmann equations as \cite{DeFelice:2011bh}
\begin{eqnarray}
   &&
2XK_{,X}-K+6X\dot{\phi}HG_{3,X}-2XG_{3,\phi}-6H^{2}G_{4}\nonumber \\
   &&
   +24H^{2}X(G_{4,X} +XG_{4,XX})-12HX\dot{\phi} G_{4,\phi X}\nonumber \\
   &&
   -6H\dot{\phi} G_{4,\phi} +2H^{3}X\dot{\phi} \left(5G_{5,X} +2XG_{5,XX}\right)\nonumber 
\\
   &&
   -6H^{2}X\left(3G_{5,\phi} +2XG_{5,\phi X}\right)=0
\,,\label{be1}
\end{eqnarray}
\begin{eqnarray}
   && K-2X(G_{3,\phi}+\ddot{\phi} 
G_{3,X})+2(3H^{2}+2\dot{H})G_{4}
\nonumber \\
   &&
   -12H^{2}XG_{4,X}-4H\dot{X}G_{4,X}-8\dot{H}
XG_{4,X}
\nonumber \\
  &&-8HX\dot{X}G_{4,XX}
+2(\ddot{\phi}+2H\dot{\phi})G_{4,\phi}+4XG_{4,\phi\phi}
\nonumber \\
   &&
   +4X(\ddot{\phi}
-2H\dot{\phi})G_{4,\phi
X}+4HX(\dot{X}-HX)G_{5,\phi
X}
\nonumber \\
&&-2X(2H^{3}\dot{\phi}+2H\dot{H}\dot{\phi}+3H^{2}\ddot{\phi})G_{5,X} 
\nonumber \\
   &&+2[2(\dot{H}X+H\dot{X})+3H^{2}X]G_{5,\phi}
   \nonumber \\
   && +4HX\dot{\phi}
G_{5,\phi\phi}
   -4H^{2}
X^{2}\ddot{\phi} G_{5,XX}=0
\,,\label{be2}
\end{eqnarray}
with dots denoting derivatives with respect to $t$, and where $H\equiv\dot{a}/a$ is the 
Hubble 
function. Additionally, variation of (\ref{Hor_L}) with respect to $\phi(t)$ gives rise 
to its 
evolution equation
\begin{equation}
\frac{1}{a^{3}}\frac{d}{dt}\left(a^{3}J\right)=P_{\phi}\,,\label{fieldeq}
\end{equation}
where
\begin{eqnarray}
J & \equiv & \dot{\phi}K_{,X}+6HXG_{3,X}-2\dot{\phi}
G_{3,\phi}-12HXG_{4,\phi X}
\nonumber
\\
 &  &
 +6H^{2}\dot{\phi}(G_{4,X}+2XG_{4,XX})\nonumber
\\
 &  & +2H^{3}X(3G_{5,X}+2XG_{5,XX})
 \nonumber
\\
 &  &-6H^{2}\dot{\phi}(G_{5,\phi}+XG_{5,\phi
X})\,,\\
P_{\phi} & \equiv & K_{,\phi}-2X\left(G_{3,\phi\phi}+\ddot{\phi}
G_{3,\phi X}\right)+6(2H^{2}+\dot{H})G_{4,\phi}\nonumber \\
 &  & +6H(\dot{X}+2HX)G_{4,\phi
X}\nonumber \\
 &  &-6H^{2}XG_{5,\phi\phi}+2H^{3}X\dot{\phi} G_{5,\phi X}\,.
 \label{Pphidef}
\end{eqnarray}
Note that in FRW geometry, $\phi$ becomes a function of $t$ only, and thus 
$X(t)=\dot{\phi}^2(t)/2$.

We proceed by examining the linear perturbations around the FRW background 
\cite{DeFelice:2011bh, Kobayashi:2011nu,Akama:2018cqv}. We work in the unitary gauge, 
i.e. 
$\delta\phi=0$, and we perturb the spatial part of the metric as $\gamma_{ij}=a^2(t) 
e^{2\zeta} (e^{h})_{ij}$, with $\zeta$ the curvature perturbation and $h_{ij}$ the tensor 
perturbation. 
We mention that the unitary 
gauge may lead to 
problems in the case where a particular quantity crosses zero at the bounce point (the 
$\gamma$-crossing of \cite{Ijjas:2017pei}), and thus one needs to 
apply the Newtonian gauge and show that the gauge variables remain non-singular, as it was 
done in  \cite{Ijjas:2017pei}. However, in our work we use the unitary gauge because 
this gauge is used in \cite{Kobayashi:2016xpl, Akama:2017jsa}  where the 
no-go theorem was presented. The fact that the ``proof'' of the no-go theorem may not be 
valid 
in the case of the $\gamma$-crossing, due to the use of the unitary gauge, could only 
serve as an additional argument against the mathematically proven universal validity of 
the no-go theorem.

Inserting these into (\ref{Hor_L}) we extract the quadratic actions for 
tensor and scalar perturbations respectively as \cite{Kobayashi:2011nu}
\begin{align}
S_h^{(2)}=\frac{1}{8}\int dt d^3x \,a^3\left[{\cal G}_T\dot h_{ij}^2
-\frac{{\cal F}_T}{a^2}(\partial h_{ij})^2\right],
\label{Sh22}
\end{align}
and 
\begin{align}
S_\zeta^{(2)}=\int dt d^3x \,a^3\left[{\cal G}_S\dot \zeta^2
-\frac{{\cal F}_S}{a^2}(\partial \zeta)^2\right].
\label{Sz22}
\end{align}
The coefficient functions are given by \cite{Kobayashi:2016xpl, Akama:2017jsa}
\begin{align}
{\cal F}_T&\equiv2 \left[ G_4-X \left(\ddot\phi G_{5,X}+G_{5,\phi}\right)\right],
\\
{\cal G}_T&\equiv2 \left[ G_4-2XG_{4,X}-X \left(H\dot\phi 
G_{5,X}-G_{5,\phi}\right)\right],
\end{align}
and 
\begin{align}
{\cal F}_S &\equiv\frac{1}{a}\frac{d\xi}{dt}-{\cal F}_T,
\label{F_Scon}
\\
{\cal G}_S &\equiv \frac{\Sigma }{\Theta^2}{\cal G}_T^2+3{\cal G}_T
\label{G_Scon},
\end{align}
 where
\begin{align}
\xi\equiv \frac{a{\cal G}_T^2}{\Theta},
\label{xi}
\end{align}
and 
 \begin{eqnarray}
\Sigma&\equiv&XK_{,X}+2X^2K_{,XX}+12H\dot\phi XG_{3,X}
\nonumber\\&&
+6H\dot\phi X^2G_{3,XX}
-2XG_{3,\phi}-2X^2G_{3,\phi X}-6H^2G_4
\nonumber\\&&
+6\Bigl[H^2\left(7XG_{4,X}+16X^2G_{4,XX}+4X^3G_{4,XXX}\right)
\nonumber\\
&&
\ \ \ \ \ \,
-H\dot\phi\left(G_{4,\phi}+5XG_{4,\phi X}+2X^2G_{4,\phi XX}\right)
\Bigr]
\nonumber\\&&
+30H^3\dot\phi XG_{5,X}+26H^3\dot\phi X^2G_{5,XX}
\nonumber\\&&
-6H^2X\bigl(6G_{5,\phi}
+9XG_{5,\phi X}+2 X^2G_{5,\phi XX}\bigr)
\nonumber\\&&
+4H^3\dot\phi X^3G_{5,XXX}\,,
\\
\Theta
&\equiv&-\dot\phi XG_{3,X}+
2HG_4-8HXG_{4,X}
\nonumber\\&&
-8HX^2G_{4,XX}+\dot\phi G_{4,\phi}+2X\dot\phi G_{4,\phi X}
\nonumber\\&&
-H^2\dot\phi\left(5XG_{5,X}+2X^2G_{5,XX}\right)
\nonumber\\&&
+2HX\left(3G_{5,\phi}+2XG_{5,\phi X}\right)\,.
\label{theta}
\end{eqnarray}
In summary, from (\ref{Sh22}) and (\ref{Sz22}) we deduce that in order for the theory to 
be free of 
ghost and gradient instabilities we must have
\begin{equation}
{\cal F}_S>0;\ {\cal G}_S>0;\ {\cal F}_T>0;\ {\cal G}_T>0.
\label{conditionss}
\end{equation}

There are two crucial assumptions for the proof of the  theoretical  {\it no-go} theorem 
\cite{Kobayashi:2016xpl}. The first is that $\Theta$ in \eqref{theta} can never cross 
zero, which implies that $\xi$ in \eqref{xi} cannot be discontinuous, which finally 
implies that ${\cal F}_S$ is a smooth function everywhere. The second (although 
not-clearly stated but definitely used) is that there is 
only one bounce, namely that the universe is always contracting before the bounce, and 
always expanding after it. Under these assumptions the proof is the following.

From the definition of ${\cal F}_S$ in \eqref{F_Scon} we deduce that the condition for 
gradient 
instabilities absence, namely ${\cal F}_S>0$, can be rewritten as
\begin{align}
\frac{d\xi}{dt}> a{\cal F}_T> 0,
\label{xibehavior}
\end{align}
which after integration from  $t_{\rm i}$ to $t_{\rm f}$ becomes
\begin{align}
 \xi_{\rm f}-\xi_{\rm i}> \int_{t_{\rm i}}^{t_{\rm f}}a{\cal F}_T dt.
\label{st2}
\end{align}

If the universe evolution is not singular one has  $a(t)>const>0$ for all times. Now, the 
integral in (\ref{st2}) for $t_{\rm f}\to\infty$ and $t_{\rm i}\to-\infty$, can be 
convergent or not, depending on the asymptotic behavior of ${\cal F}_T$. In the case where 
it is non-convergent relation (\ref{st2}) implies
$
-\xi_{\rm f}<-\xi_i-\int_{t_{\rm i}}^{t_{\rm f}}a{\cal F}_Tdt,
$
and since the integral is a positive and increasing function of $t_{\rm f}$ (${\cal 
F}_T>0$ 
according to (\ref{conditionss})), for sufficiently large 
$t_{\rm f}$ the right hand side will become negative. This means that $\xi_{\rm f}>0$. On 
the other 
hand writing (\ref{st2}) as
$
-\xi_{\rm i}>-\xi_{\rm f}+\int_{t_{\rm i}}^{t_{\rm f}}a{\cal F}_Tdt 
$
we see that for $t_{\rm i}\to -\infty$ the right hand side will become positive and thus 
$\xi_{\rm 
i}<0$. Hence, since $\xi_{\rm f}>0$ and $\xi_{\rm i}<0$ one could deduce that $\xi$ 
crosses zero. 
However, according to (\ref{xi}), if $\xi$ is not discontinuous then it can never cross 
zero for a nonsingular bounce, namely for $a(t)>const>0$ (note that ${\cal 
G}_T^2>0$ for every theory 
that has general relativity as a particular limit, since in general relativity $G_4=1$). 
Hence,  in \cite{Kobayashi:2016xpl} it is concluded that the nonsingular 
condition $a(t)>const>0$ must be relaxed if we desire not to have instabilities (i.e. if 
${\cal F}_T>0$), and thus $a(t)$ should be zero at a specific time. Finally, the proof is 
completed by considering the case where the integral in (\ref{st2}) is convergent, which 
requires ${\cal F}_T\rightarrow0$ sufficiently fast either in the asymptotic past or 
future. However, as ${\cal F}_T\rightarrow0$ the normalization of vacuum quantum 
fluctuations implies that they diverge (strong-gravity problem), and thus tensor 
perturbations will 
asymptotically exhibit pathologies. 

In summary, under the assumption that $\Theta$ in \eqref{theta} can never cross zero, 
i.e. that $\xi$ in \eqref{xi} cannot be discontinuous, and that there is only one bounce, 
in \cite{Kobayashi:2016xpl} it was shown that the condition for instabilities 
absence 
in the tensor sector, namely ${\cal F}_T>0$, 
implies that $a(t)$ should be zero at a specific time, and hence a nonsingular bounce is 
impossible in the framework of Horndeski/Galileon cosmology. Finally, one can extend the 
above arguments and proof in the case where there are more degrees of freedom in the 
scalar perturbations \cite{Kobayashi:2016xpl}, as well as in the case of multi-galileon 
theory \cite{Akama:2017jsa}.

\section{Evading the theoretical no-go theorem}
\label{Evading}

In the previous section we reviewed the  theoretical  {\it no-go} theorem presented 
in \cite{Kobayashi:2016xpl}, stating that a nonsingular bounce cannot be 
realized in Horndeski/Galileon cosmology if we desire not to have ghost and gradient 
instabilities. As we mentioned, the proof is based on two very strong assumptions, namely 
that $\Theta$ in \eqref{theta} can never cross zero and hence that $\xi$ in \eqref{xi} 
cannot be discontinuous, and that there is only one bounce. However, as we will show in 
this section, not only these assumptions can be violated in usual bouncing scenarios, 
but on the contrary for general sub-classes of the theory it is impossible not to violate 
them.

The main condition of the bounce realization is that the Hubble function must be zero at 
the bounce point. Thus, as one can see, the majority of terms in $\Theta$ definition in 
\eqref{theta} become zero at a general bounce. Now, observing the first Friedmann equation 
of Horndeski/Galileon cosmology, namely Eq. (\ref{be1}), we can see that if the function 
$K(\phi,X)$ becomes 
zero at $X=0$, then the above main bounce condition is realized if $X$, i.e. 
$\dot{\phi}$, 
becomes zero at the bounce point. But $\dot{\phi}=0$ implies that $\Theta$ in 
\eqref{theta} crosses zero at the bounce point, or equivalently  $\xi$ in \eqref{xi} 
becomes discontinuous. Hence, we 
conclude that the assumption on which the  theoretical  {\it no-go} theorem is based is 
always violated in a nonsingular bounce if $K(\phi,0)=0$. Note that $K(\phi,0)=0$ 
(which for instance is satisfied in the ``kinetic'' choices where $K$ is a polynomial of 
$X$ \cite{Easson:2011zy}) is a sufficient condition, not a necessary one, since $\Theta$ 
can become 
zero at the bounce point for other suitable choices of $K(\phi,X)$ too. However, 
the above sub-case ensures the successful evading of the above  theoretical  {\it no-go} 
theorem.
\begin{figure}[ht]
\centering
\includegraphics[width=7.cm,height=4.cm]{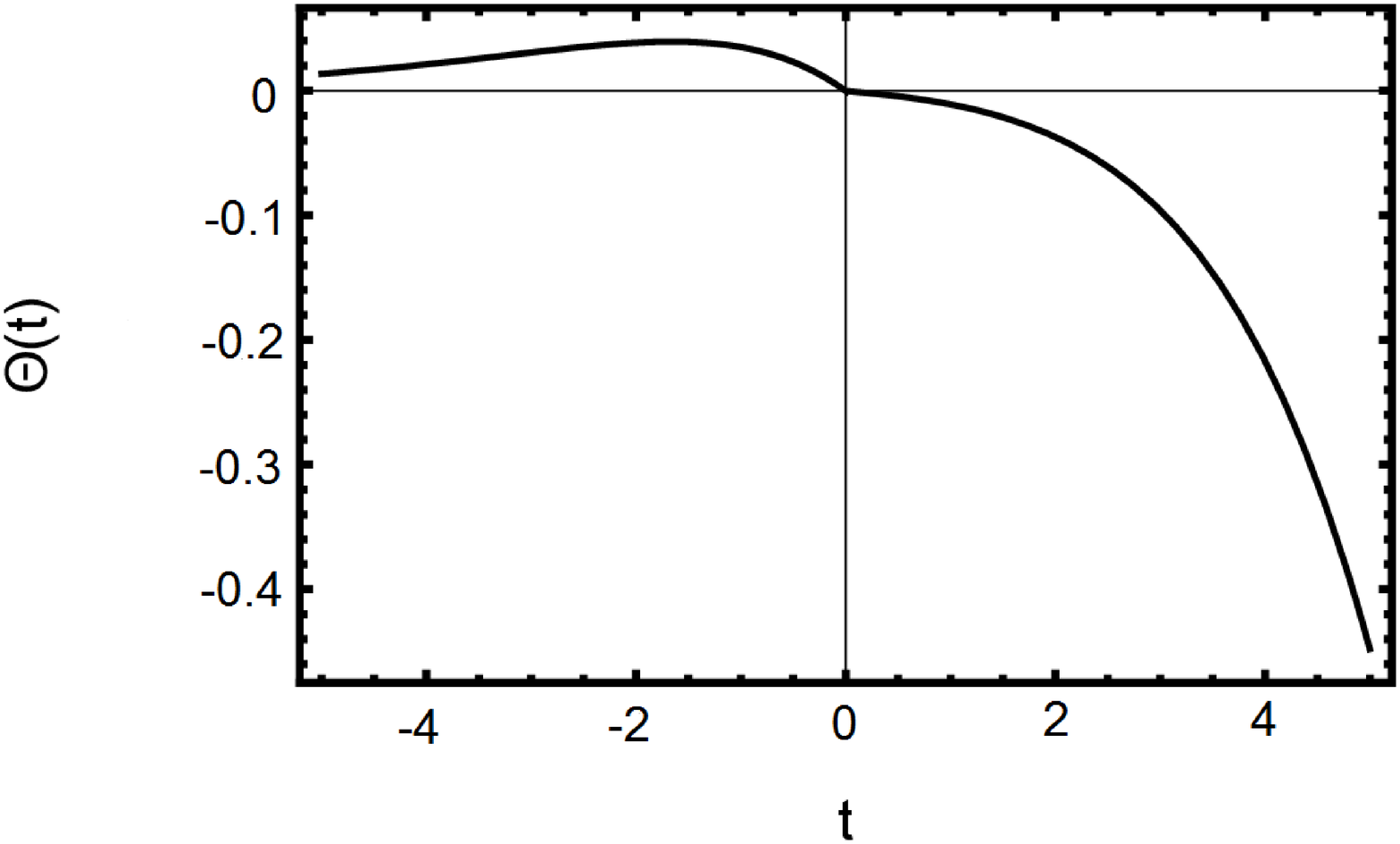}\\
\includegraphics[width=7.cm,height=4.cm]{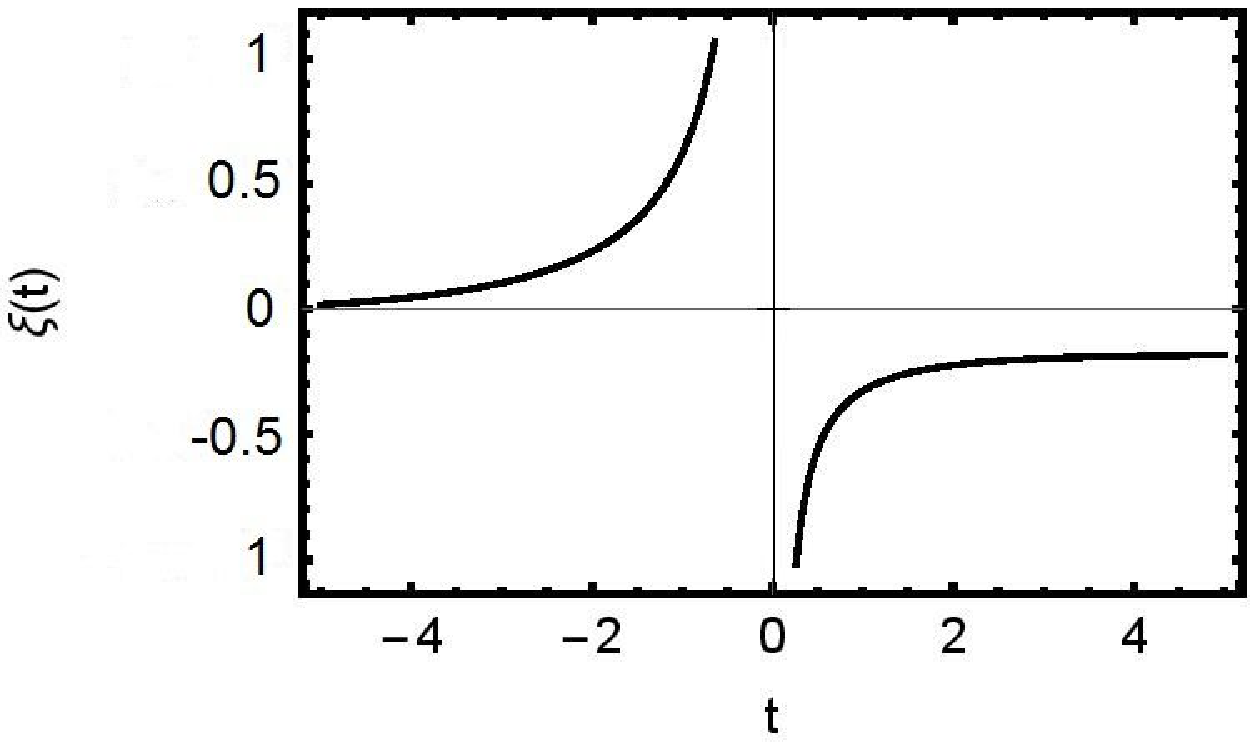}
\caption{\small {\em 
The evolution of the functions $\Theta(t)$ (upper graph) and $\xi(t)$ (lower graph), for 
the 
nonsingular bounce (\ref{bounceform}) with $a_b=0.2$, $B=10^{-5}$, under the choice 
$K=X^2$, $G_{4}
=1+X^2$, $G_5=0$. All quantities are measured in units where $M_{pl}=1$, and the vertical 
line at $t=0$ is drawn for convenience. }}
\label{xithetafig}
\end{figure}

Let us provide a specific example where the  theoretical  {\it no-go} theorem is 
evaded as we 
described, and a nonsingular bounce free from ghost and gradient instabilities can be 
realized in 
the framework of Horndeski/Galileon cosmology.
We will follow the  method presented in \cite{Banerjee:2016hom}, in which 
one inserts the desired scale factor, as well as the ansatzes of some of the involved 
functions, and reconstructs the rest of them in order to obtain self-consistency.
We first consider a specific nonsingular 
bounce 
scale factor of the form 
\begin{equation}
a(t)=a_b(1+Bt^2)^{1/3},
\label{bounceform}
\end{equation}
with $a_b$ the scale factor value at the bounce and $B$ a positive parameter, i.e. time 
varies 
between $-\infty$ and $+\infty$ and the bounce is realized at $t=0$. Additionally, we 
consider a 
shift-symmetric Horndeski/Galileon model with 
\begin{equation}
 K=X^2, \ \ G_{4}=1+X^2,\ \  G_5=0.
\label{ourchoice}
\end{equation}
Thus, inserting these into the Friedmann equations and assuming that 
$G_{3}(\phi,X)=G_3(X)$ one can 
numerically extract the solution for $\phi(t)$ and reconstruct the $G_3(X)$ form that 
generates the 
above bounce realization \cite{Banerjee:2016hom}. Finally, knowing the behaviour of all 
background 
quantities, we can numerically calculate the perturbation quantities ${\cal F}_S$, ${\cal 
G}_S$, ${\cal F}_T$, ${\cal G}_T$ and examine whether they are positive, i.e. satisfying 
the conditions for 
absence of ghost and gradient instabilities (\ref{conditionss}).

\begin{figure}[ht]
\centering
\includegraphics[width=6.cm,height=3.8cm]{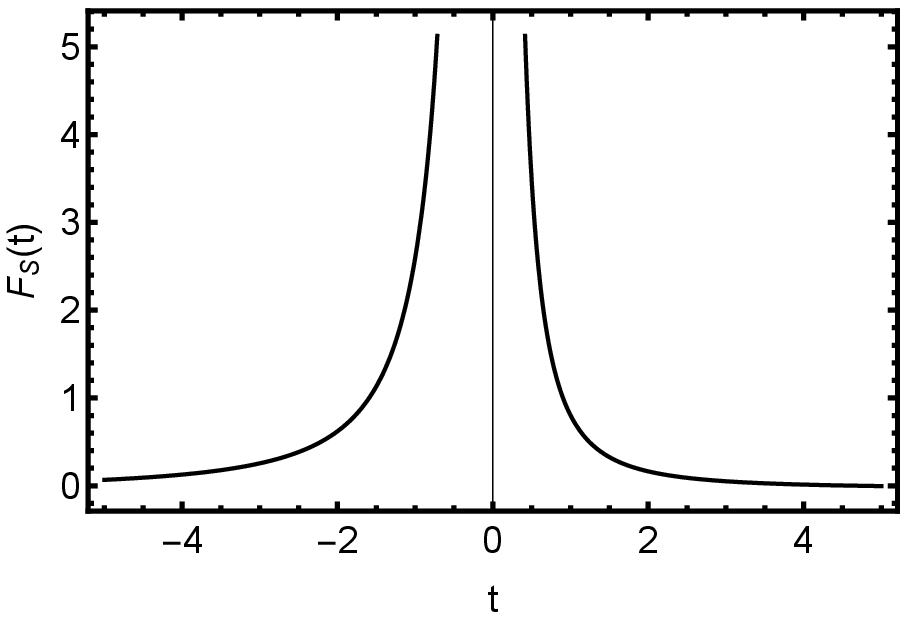}\\
\includegraphics[width=6.cm,height=3.8cm]{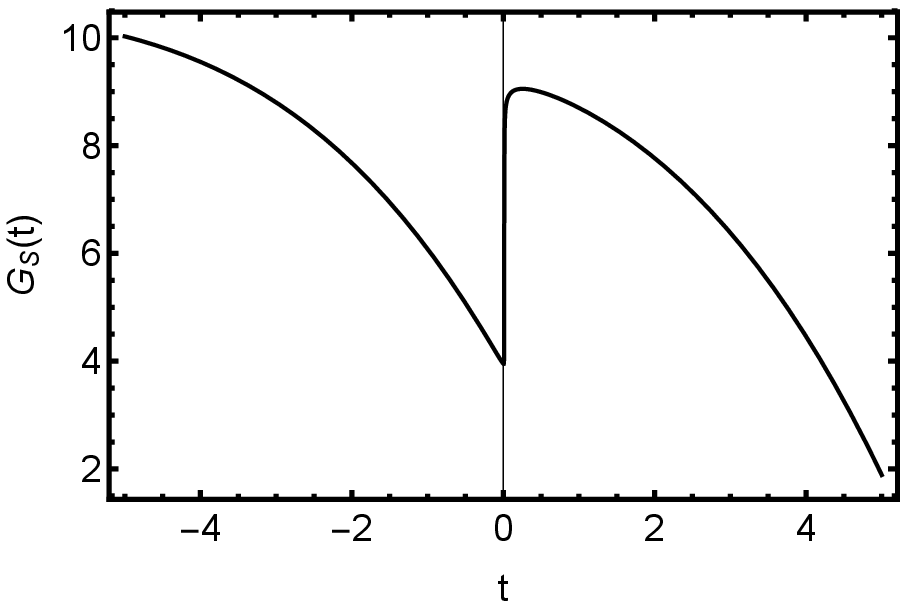}\\
\caption{\small {\em  
The evolution of the quantities ${\cal F}_S(t)$ (upper graph) and ${\cal G}_S(t)$ (lower 
graph) 
related to scalar perturbations, for the nonsingular bounce (\ref{bounceform}) with 
$a_b=0.2$, 
$B=10^{-5}$, under the choice $K=X^2$, $G_{4}=1+X^2$, $G_5=0$. All quantities are 
measured in units where $M_{pl}=1$, and the vertical 
line at $t=0$ is drawn for convenience.
}}
\label{FSGSfig}
\end{figure}
\begin{figure}[!]
\centering
\includegraphics[width=6.cm,height=3.8cm]{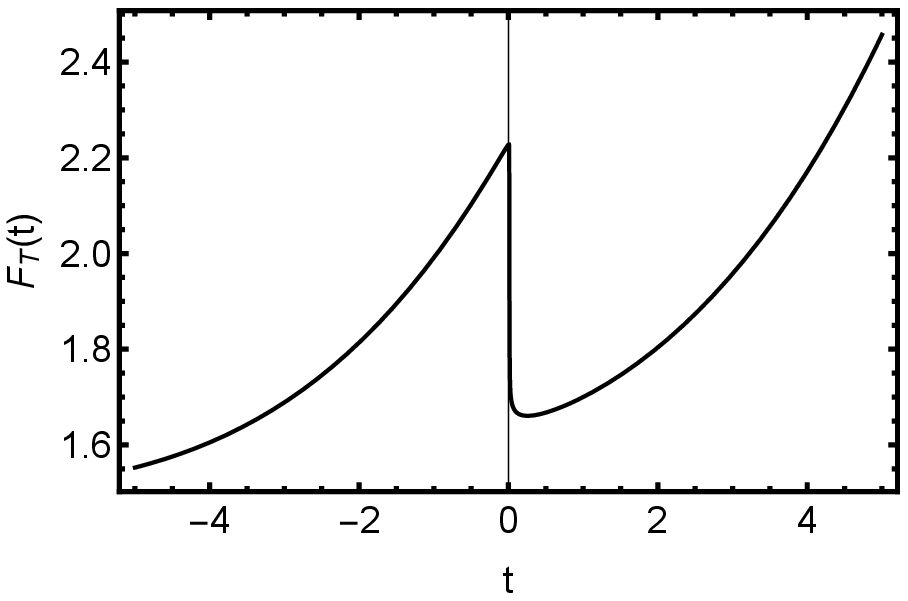}\\
\includegraphics[width=6.cm,height=3.8cm]{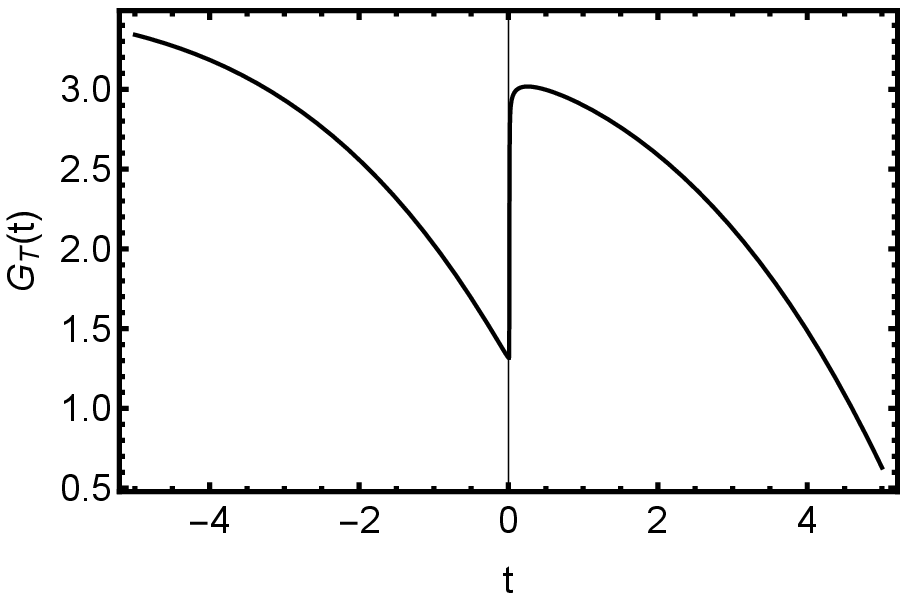}
\caption{\small {\em 
The evolution of the quantities  ${\cal F}_T(t)$ (upper graph) and ${\cal G}_T(t)$ (lower 
graph) 
related to tensor perturbations, for the nonsingular bounce (\ref{bounceform}) with $ 
a_b=0.2$, 
$B=10^{-5}$, under the choice $K=X^2$, $G_{4}=1+X^2$, $G_5=0$. All quantities are 
measured 
in units where $M_{pl}=1$, and the vertical 
line at $t=0$ is drawn for convenience.
}}
\label{FTGTfig}
\end{figure}
In Fig. \ref{xithetafig} we depict the behavior of  $\Theta(t)$ and $\xi(t)$ for the 
nonsingular bounce (\ref{bounceform}). As we can see, the basic assumption of the  
theoretical  {\it no-go}  theorem is evaded, namely $\Theta(t)$ crosses zero at the 
bounce point, and thus $\xi(t)$ becomes discontinuous and transits from positive to 
negative values without crossing zero and being always an increasing function. 
Additionally, in Fig. \ref{FSGSfig} we present the corresponding behavior of the 
quantities ${\cal F}_S$ and ${\cal G}_S$ that are related to  scalar perturbations, while 
in Fig. \ref{FTGTfig} we show the corresponding behavior of ${\cal F}_T$ and ${\cal G}_T$ 
that are related to  tensor perturbations. As we observe all of them are positive and thus 
the conditions (\ref{conditionss}) for the absence of ghost and gradient instabilities are 
satisfied.

In summary, with the general justification we presented in the beginning of this 
section, we showed that a nonsingular bounce free from ghost and gradient instabilities 
can indeed be realized in the framework of Horndeski/Galileon cosmology, and without 
loss of generality we verified it with the specific example given above. 
 
We continue the investigation examining some ``improvements'' of the theoretical  no-go 
theorem that have appeared in the literature. In \cite{Akama:2017jsa} it was argued that 
the no-go theorem could also be proved in the case where  $\Theta(t)$ crosses zero, i.e 
$\xi(t)$ becomes discontinuous, at the bounce point, however again under the crucial 
assumption that this happens only one time (which due to the fact that $\xi(t)$ must be 
monotonous according to (\ref{xibehavior}) allows one to deduce that $\lim_{t\rightarrow 
\pm\infty}\xi=const.$). Note that similar arguments under the single-bounce assumption 
are also made in \cite{Mironov:2018oec}, where $\Theta$ is denoted by $\gamma$, and thus 
the $\Theta$-crossing is called $\gamma$-crossing (nevertheless even assuming a single 
bounce these authors do not exclude the evading of the no-go theorem in the case where  
$\Theta$ (i.e. $\gamma$) and $G_T$, namely the denominator and numerator in (\ref{xi}),  
vanish at the same time). 
 
As we mentioned, the assumption of a single bounce remains crucial in the updated 
versions of the theoretical no-go theorem  \cite{Kobayashi:2016xpl, Akama:2017jsa, 
Kolevatov:2016ppi, Kolevatov:2017voe} (see also 
\cite{Libanov:2016kfc,Mironov:2018oec,Evseev:2017jek}), since the proof 
does admit that the nonsingular bounce itself can indeed be free of any pathologies, 
however suitably far from the bounce, either in the infinite past or in the infinite 
future, even if ${\cal F}_S$, ${\cal G}_S$, ${\cal F}_T$, ${\cal G}_T$ remain 
non-negative we will have  ${\cal F}_T\rightarrow0$ (or  ${\cal F}_S\rightarrow0$) which 
leads to pathologies and the onset of strong coupling. Although the principle that in 
order to study a local bounce one should examine the global behavior of the universe is a 
bit uncomfortable\footnote{This issue, namely whether a pathologies-free bounce that may 
be accompanied by pathologies in the phase far before or far  after the bounce is 
acceptable or not, has led to a debate in the literature 
\cite{Ijjas:2016tpn,Ijjas:2016vtq,Dobre:2017pnt}.} (we mention that for instance in the 
specific example we presented 
above  the time scale of the evolution depends on the parameter $B$, and thus choosing it 
arbitrarily small could push the ${\cal 
F}_T\rightarrow0$, ${\cal 
F}_S\rightarrow0$  regimes arbitrarily far), 
still the assumption that the universe expands forever before or after the bounce is a 
very strong one.
 
Indeed, it is known that many modified gravities may lead to cyclic cosmology 
\cite{Lehners:2008vx,Cai:2011bs}, namely to an infinite series of bounces and 
turnarounds, and Horndeski/Galileon theory is one of them 
\cite{Leon:2012mt,Banerjee:2016hom,Saridakis:2018fth}. Hence, one can 
clearly see that in a multiple realization of the pathologies-free nonsingular bounce, 
which the proof of the theoretical no-go theorem does admit that it can exist, the 
universe never reaches the regime  ${\cal 
F}_T\rightarrow0$ and/or ${\cal 
F}_S\rightarrow0$, since there is not infinite past and infinite future regime before and 
after any bounce respectively.
 
\begin{figure}[ht]
\centering
\includegraphics[width=6.cm,height=4.cm]{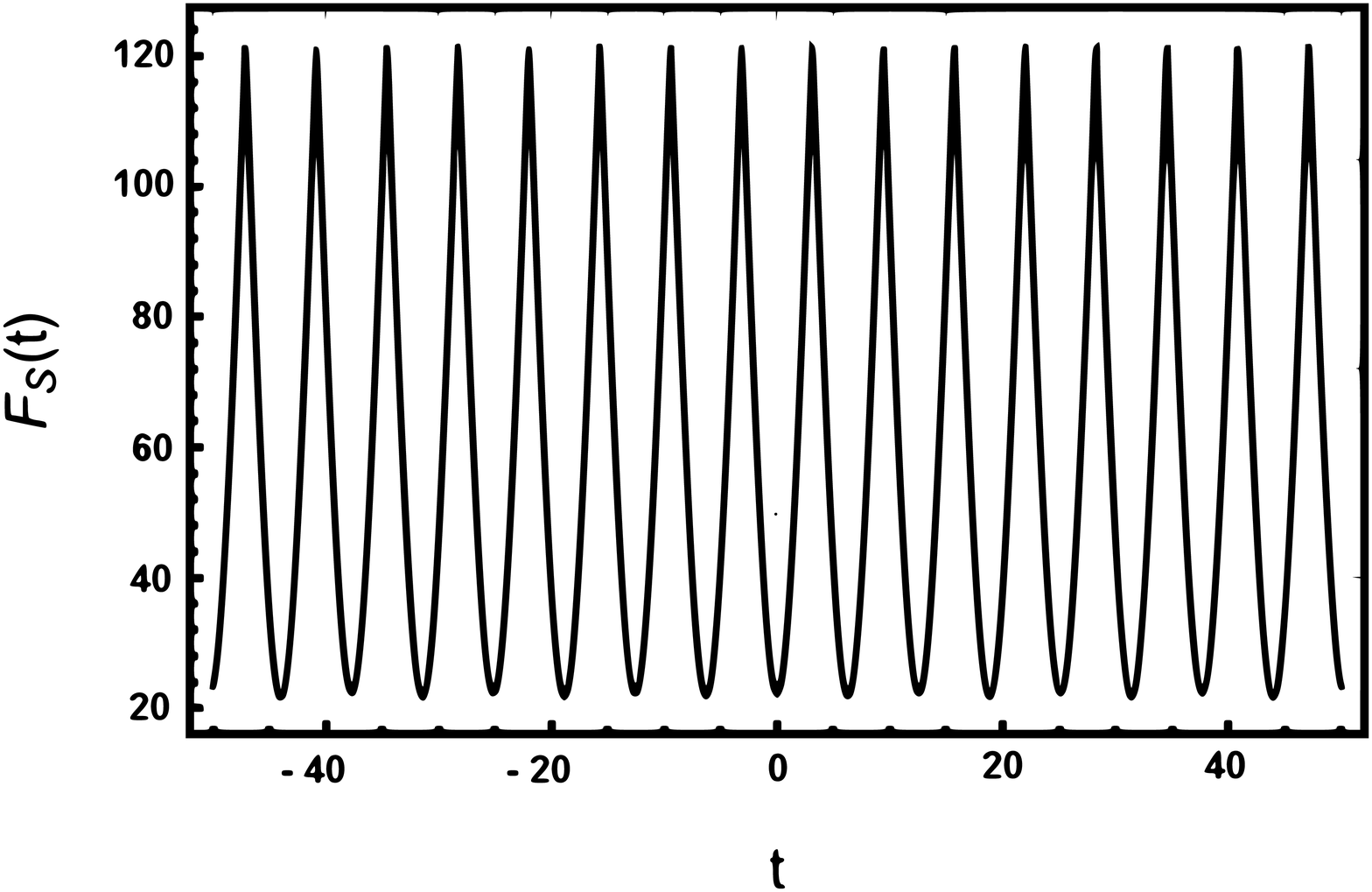}\\
\includegraphics[width=6.cm,height=4.cm]{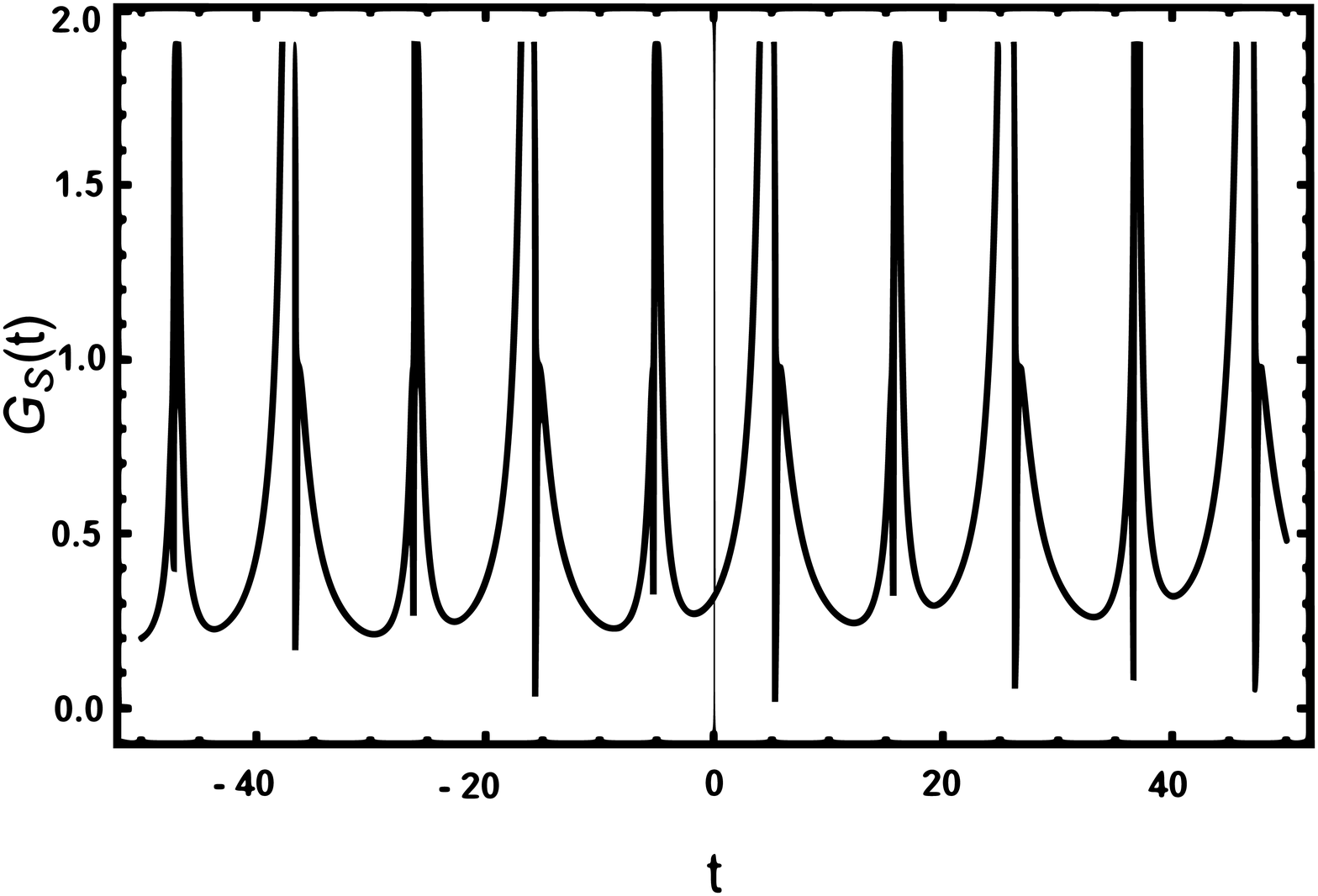}\\
\caption{\small {\em  
The evolution of the quantities ${\cal F}_S(t)$ (upper graph) and ${\cal G}_S(t)$ (lower 
graph) 
related to scalar perturbations, for the cyclic scale factor (\ref{cyclicscaefactor}) 
with $a_c=0.01$, $A=10^{-4}$, $\omega=0.5$, under the choice $K=X+V(\phi)$, $G_3=X$, 
$G_{4}=1+X^2$, $G_5=0$.
All quantities are 
measured in units where $M_{pl}=1$, and the vertical 
line at $t=0$ is drawn for convenience.
}}
\label{FSGSfigcycl}
\end{figure}

\begin{figure}[!]
\centering
\includegraphics[width=6.cm,height=4.cm]{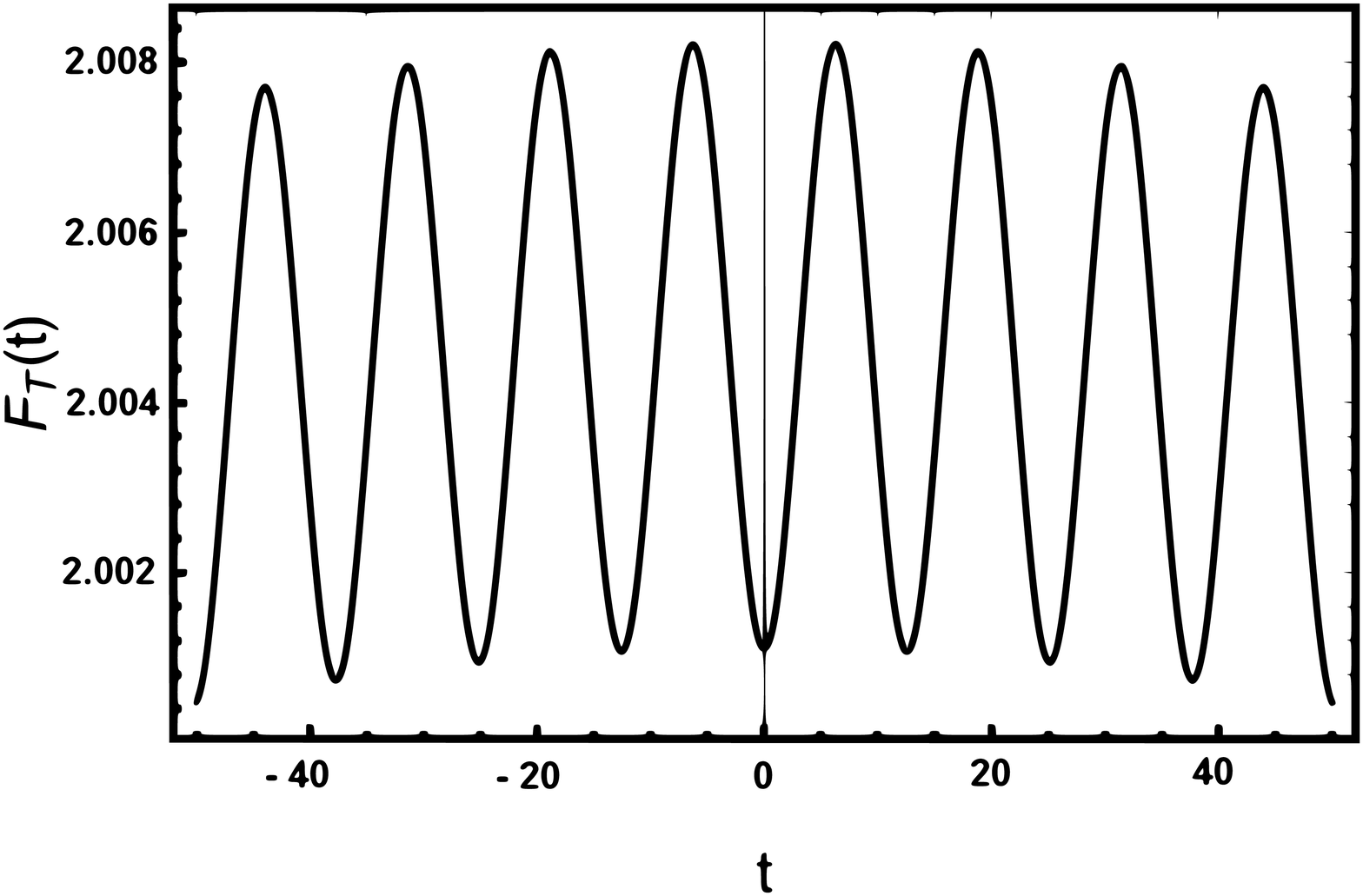}\\
\includegraphics[width=6.cm,height=4.cm]{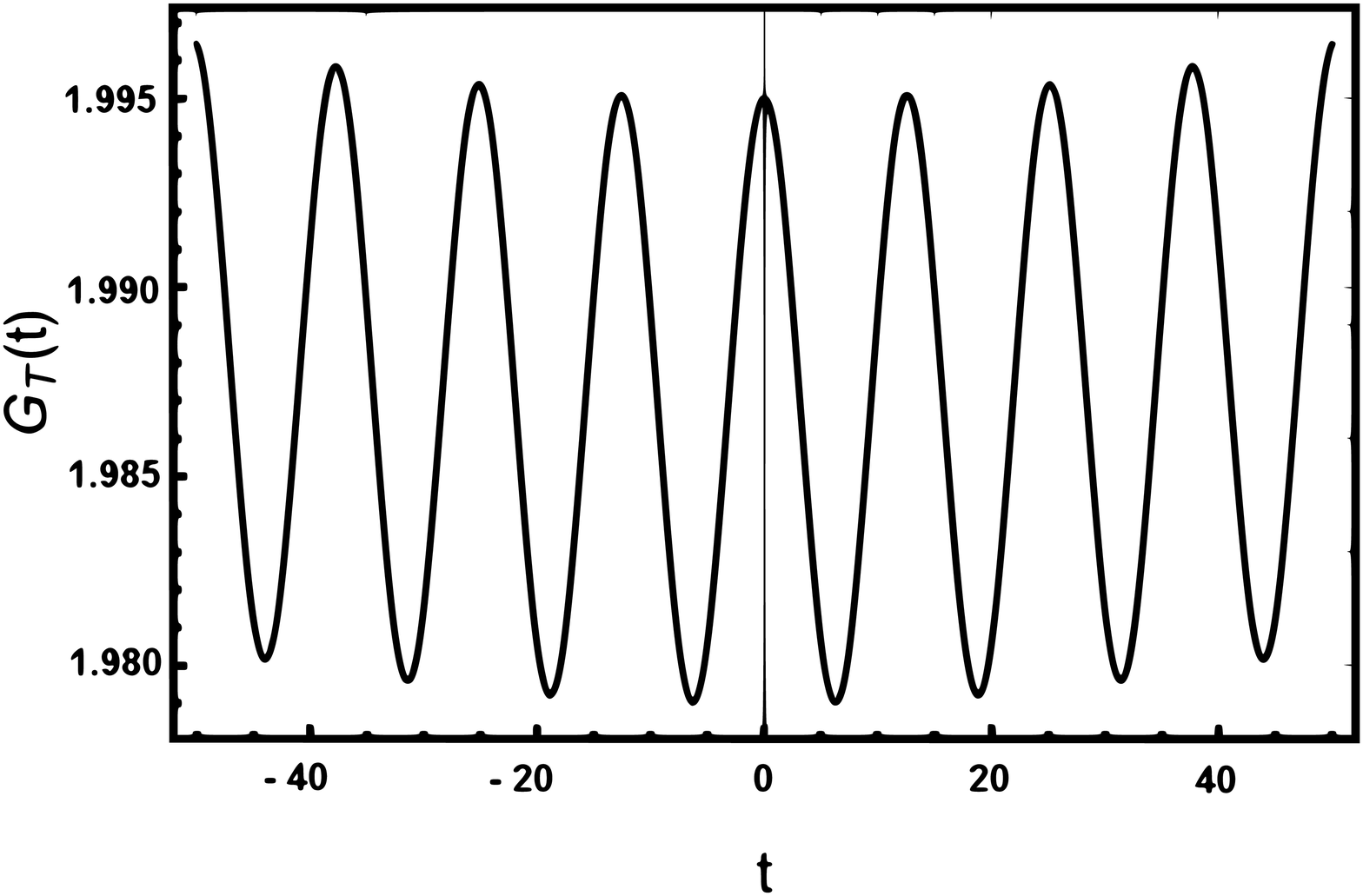}
\caption{\small {\em 
The evolution of the quantities  ${\cal F}_T(t)$ (upper graph) and ${\cal G}_T(t)$ (lower 
graph) 
related to tensor perturbations,  for the cyclic scale factor (\ref{cyclicscaefactor}) 
with $a_c=0.01$, $A=10^{-4}$, $\omega=0.5$, under the choice $K=X+V(\phi)$, $G_3=X$, 
$G_{4}=1+X^2$, $G_5=0$.
All quantities are 
measured in units where $M_{pl}=1$, and the vertical 
line at $t=0$ is drawn for convenience.
}}
\label{FTGTfigcycl}
\end{figure}

In order to again give a specific example of such a possibility  we follow the 
method of  \cite{Banerjee:2016hom}  and we impose the nonsingular oscillating scale 
factor \footnote{Cyclic cosmology may   exhibit 
the  old entropy problem (although the works of 
Frampton et. al. may offer ways to evade it, see e.g. \cite{Baum:2007de}), nevertheless 
this is a 
completely different issue from the mathematical ``no-go theorem''.}
 \begin{equation}
 \label{cyclicscaefactor}
 a(t)=A \sin(\omega t)+a_{c},
 \end{equation}
  where  $a_{c}-A>0$ is the scale factor value at the bounce, with  $A+a_c$ the scale 
factor value at the turnaround. Note that this is not the most general cyclic 
scale factor, since its minima and maxima happen at the same values, however it is 
adequate for the subsequent discussion.
We moreover consider  $K=X+V(\phi)$, $G_3=X$, 
$G_{4}=1+X^2$, $G_5=0$, while we must also include the matter sector in 
order to be consistent with the whole universe history (the matter sector   
does not interfere with the discussion on the bounce stability and the no-go theorem).
Inserting these into the Friedmann equations we can 
numerically extract the solution for $\phi(t)$ and reconstruct the $V(\phi)$ form that 
generates the above cyclic scale factor \cite{Banerjee:2016hom}. Finally, knowing the 
behaviour of all background quantities, we can numerically calculate   ${\cal F}_S$, 
${\cal G}_S$, ${\cal F}_T$, ${\cal G}_T$. In  Fig. \ref{FSGSfigcycl} we present the 
evolution of ${\cal F}_S$ 
and ${\cal G}_S$ that are related to  scalar perturbations, while in Fig. 
\ref{FTGTfigcycl} 
we show the corresponding behavior of ${\cal F}_T$ and ${\cal G}_T$ that are related to 
 tensor perturbations. As we observe all of them are positive and thus the conditions 
(\ref{conditionss}) for the absence of ghost and gradient instabilities are satisfied. 
Furthermore, the regimes  ${\cal 
F}_T\rightarrow0$ and/or ${\cal 
F}_S\rightarrow0$ are never reached since after any bounce the universe cannot expand 
forever since it is followed by a turnaround and a next bounce.

Hence, as we showed in detail in this section,  a nonsingular bounce free from ghost and 
gradient instabilities can indeed be realized in the framework of Horndeski/Galileon 
cosmology. The reason behind the evading of the theoretical no-go theorem is that  
$\Theta(t)$ crosses zero at the 
bounce point, and thus $\xi(t)$ becomes discontinuous and transits from positive to 
negative values without crossing zero and being always an increasing function. 
Additionally, even if one ``improves'' the no-go theorem by claiming that although a 
nonsingular bounce free of pathologies can be realized at some point, it will lead to 
strong-gravity-related pathologies at infinite past or infinite future, this can also be 
evaded by considering cyclic cosmology, namely an infinite sequence of  nonsingular 
bounces free of pathologies, which forbids the universe to reach the ``problematic'' 
regime at infinite past or infinite future. Lastly, note also the interesting 
possibility that a nonsingular bounce free of pathologies is accompanied by a singular 
bounce  free of pathologies, in which case all the arguments of the theoretical no-go 
theorem of  \cite{Kobayashi:2016xpl, Akama:2017jsa, 
Kolevatov:2016ppi, Kolevatov:2017voe,Libanov:2016kfc,Mironov:2018oec} collapse, and the 
nonsingular bounce free of pathologies can clearly exist.

\section{Conclusions}
\label{Conclusions}

In this work we showed that a nonsingular bounce, free of ghosts and gradient 
instabilities, can be realized in the framework of Horndeski or generalized Galileon 
cosmology. This result was known through specific models \cite{Easson:2011zy, Cai:2012va, 
Battarra:2014tga, Qiu:2015nha, Banerjee:2016hom, Ijjas:2016tpn, Ijjas:2016vtq, 
Ijjas:2017pei, Saridakis:2018fth}, however in this work we proved why the  theoretical  
{\it no-go} theorem which states that such a realization is impossible 
\cite{Kobayashi:2016xpl,Akama:2017jsa} can be evaded. 

In particular, we first reviewed 
that this  theoretical  {\it no-go} theorem is based on two   very strong assumptions, 
namely  that a 
particular quantity, $\xi$ in \eqref{xi}, cannot be discontinuous, and that there is 
only one bounce. Concerning the first assumption  we showed 
that not only   can be violated   in a general 
Horndeski/Galileon scenario, but that it is necessarily violated at the bounce point in 
the subclass of Horndeski/Galileon gravity in which 
$K(\phi,0)=0$ (as for instance in the kinetic choices where $K$ is a polynomial of $X$). 
In order to make the analysis more transparent, and without loss of generality, 
we provided an explicit example where a nonsingular bounce is realized, with all 
stability conditions being satisfied. Concerning the second assumption, which is also 
crucial in improved versions of the theoretical no-go theorem which claim that even if a 
nonsingular bounce free of pathologies can be realized it will lead to pathologies in the 
infinite past or infinite future, we showed that it can be evaded by considering cyclic 
cosmology, with an infinite sequence of  nonsingular bounces free of pathologies, which 
forbids the universe to reach to the ``problematic'' regime at infinite past or infinite 
future. In this case we also provided a specific example with the above behavior, with 
all stability conditions being satisfied eternally.

In conclusion, stable nonsingular bounce realizations are not mathematically impossible 
in Horndeski/Galileon 
cosmology, which may serve as an additional advantage for this class of gravitational 
modification.

\section*{Acknowledgments}
We are grateful to  J. Barrow, Y. Cai, D. Easson, X. Gao, T. Kobayashi, S. Mironov, T. 
Qiu,  A. 
Vikman, D. G. Wang, P. Zhang and M. Zhu for 
stimulating discussions.
This article is based upon work from COST Action ``Cosmology and Astrophysics Network for 
Theoretical Advances and Training Actions'', supported by COST (European Cooperation in 
Science and Technology).
The work of YFC is supported in part by the National Youth Thousand Talents Program of 
China, by the NSFC (Nos. 11722327, 11653002, 11421303, J1310021), by the CAST Young Elite 
Scientists Sponsorship (2016QNRC001), and by the Fundamental Research Funds for Central 
Universities. The work of ENS  is partly supported  by the International Visiting 
Professorship at USTC. Part of numerics are operated on the computer cluster LINDA in the 
particle cosmology 
group at USTC.

\end{document}